\documentclass[lettersize,journal]{IEEEtran}
\IEEEoverridecommandlockouts
 \usepackage{color}
 \usepackage{tabularray}
\usepackage[table,xcdraw]{xcolor}
\usepackage{dblfloatfix}
\usepackage{cite}
\usepackage{multicol}
\usepackage{diagbox}
\usepackage{tabularray}
\usepackage{amssymb}
\usepackage{amsfonts}
\usepackage{amsmath}
\usepackage[T1]{fontenc}
\usepackage[utf8]{inputenc}
\usepackage[english]{babel}
\usepackage{pifont}
\usepackage{booktabs}
\usepackage{amsmath}
\usepackage{mathtools}
\usepackage{ellipsis}
\usepackage{textcomp}
\usepackage{cite}
\usepackage[pdftex]{graphicx}
\usepackage{subfigure}
\usepackage{subcaption}
\usepackage{epsfig}
\usepackage{epstopdf}
\usepackage{caption}
\definecolor{Mycolor1}{HTML}{ff0000}
\definecolor{Mycolor2}{HTML}{13beb8}
\definecolor{Mycolor3}{HTML}{ffea00}
\definecolor{Mycolor4}{HTML}{4dd0e1}
\definecolor{Mycolor5}{HTML}{ab47bc}
\definecolor{Mycolor6}{HTML}{ffa726}
\definecolor{Mycolor7}{HTML}{bcaaa4}
\definecolor{Mycolor8}{HTML}{43a047}
\definecolor{Mycolor9}{HTML}{81d4fa}
\definecolor{Mycolor10}{HTML}{999900}
\definecolor{Mycolor11}{HTML}{FFEBCC}
\definecolor{Mycolor12}{HTML}{D1E9E2}
\usepackage{algorithm}
\definecolor{customblue}{rgb}{0.30, 0.65, 0.89} 
\definecolor{customorange}{rgb}{1.00, 0.65, 0.20} 
\usepackage{fixltx2e}
\usepackage{algpseudocode}
\usepackage{epstopdf}
\usepackage{psfrag}
\usepackage{grffile}
\usepackage{amsthm}

\usepackage{caption}

\theoremstyle{definition}

\theoremstyle{remark}

\theoremstyle{plain}

\newcommand{\RNum}[1]{\uppercase\expandafter{\romannumeral #1\relax}}
\usepackage{setspace}

\def\BibTeX{{\rm B\kern-.05em{\sc i\kern-.025em b}\kern-.08em
    T\kern-.1667em\lower.7ex\hbox{E}\kern-.125emX}}
\begin{document}
\title{Wireless MAC Protocol Synthesis and Optimization with Multi-Agent Distributed Reinforcement Learning
}
\author{Navid~Keshtiarast,
        ~Oliver~Renaldi,
and~Marina~Petrova,~\IEEEmembership{Senior~Member,~IEEE}
	\thanks{Navid~Keshtiarast, Oliver~Renaldi and  Marina~Petrova are with Mobile Communications and Computing Group, at RWTH Aachen University, Germany (e-mail: \{navid.keshtiarast@mcc, oliver.renaldi, petrova@mcc\}.rwth-aachen.de)}
}
\maketitle
\begin{abstract}
 In this letter, we propose a novel Multi-Agent Deep Reinforcement Learning (MADRL) framework for Medium Access Control (MAC) protocol design. Unlike centralized approaches, which rely on a single entity for decision-making, MADRL empowers individual network nodes to autonomously learn and optimize their MAC based on local observations. Leveraging ns3-ai and RLlib, as far as we are aware of, our framework is the first of a kind that enables distributed multi-agent learning within the \mbox{ns-3} environment, facilitating the design and synthesis of adaptive MAC protocols tailored to specific environmental conditions. We demonstrate the effectiveness of the MADRL MAC framework through extensive simulations, showcasing superior performance compared to legacy protocols across diverse scenarios. Our findings highlight the potential of MADRL-based MAC protocols to significantly enhance Quality of Service (QoS) requirements for future wireless applications. 
\end{abstract}
 \vspace{-0.15cm}
\section{Introduction}
\label{sec:Introduction}
Wireless networks are continuously faced with a multitude of demands, ranging from high-reliability and low-latency connectivity to support of bandwidth-intensive applications such as virtual reality (VR), gaming, and holographic video. These various applications highlight the growing need for adaptable and application-specific channel access mechanisms and resource allocation. However, the large number of configurable parameters and their entangled inter-dependencies in the medium access control (MAC) layer, for example, pose challenges to optimize and fine-tune protocols with traditional methods due to the dynamic nature of wireless networks \cite{Naderializadeh}, especially in uncoordinated environments such as unlicensed bands lacking a centralized authority to regulate channel access across all network nodes.

In recent years, AI-driven approaches, particularly deep reinforcement learning (DRL), have shown great promise in optimizing wireless network performance \cite{Falko} by letting MAC protocols to learn and adapt autonomously based on real-time feedback from the environment, and ensuring more intelligent and adaptive behavior. However, most of the existing works in this area optimize and configure only a few parameters in the MAC layer or physical layer\cite{9946841}. Some recent studies also explore applications of DRL for generating protocols and signalling for cellular  \cite{Mehdi,Jakob1} and Wi-Fi networks \cite{keshtiarast, Pasandi}. While these studies make significant step forward, the proposed solutions rely on centralized entities for training. Even in instances where distributed inference is employed, the overarching approach remains centralized, hampering scalability and adaptability, particularly in dynamic network environments where decentralized decision-making is desirable.
Expanding upon our prior work, \cite{keshtiarast, Peng}, which proposed a DRL-based MAC design framework with centralized learning and execution mechanisms for Wi-Fi networks, in this letter, we introduce a Multi-Agent Deep Reinforcement Learning
(MADRL) framework that incorporates both centralized and distributed learning, along with distributed inference. This advancement increases flexibility and adaptability, empowering individual nodes to manage diverse traffic loads and environmental conditions effectively. We implement our MADRL framework within \mbox{ns-3} by integrating ns3-ai\cite{ns-3ai} and RLlib.  To the best of our knowledge, this is the first implementation that enables distributed multi-agent learning using ns3-ai and Ray RLlib in \mbox{ns-3} environment. Furthermore, we extend the capabilities of our framework to support 5G New Radio Unlicensed (5G \mbox{NR-U}) technology, which demonstrates the versatility of our solution across emerging wireless technologies beyond Wi-Fi networks. We test and verify the performance of our framework and the learnt protocols through extensive system-level
simulations and demonstrate the superior performance of the MADRL-synthesized protocols over legacy protocols across diverse scenarios. This underlines the potential of our framework to significantly
enhance QoS for future applications through a novel protocol design approach.

\section{System Concept} 
\begin{figure}[h!]
\vspace{-0.29cm}\centering\includegraphics[width=0.43\textwidth,trim = 43mm 51.5mm 52mm 38mm,clip]{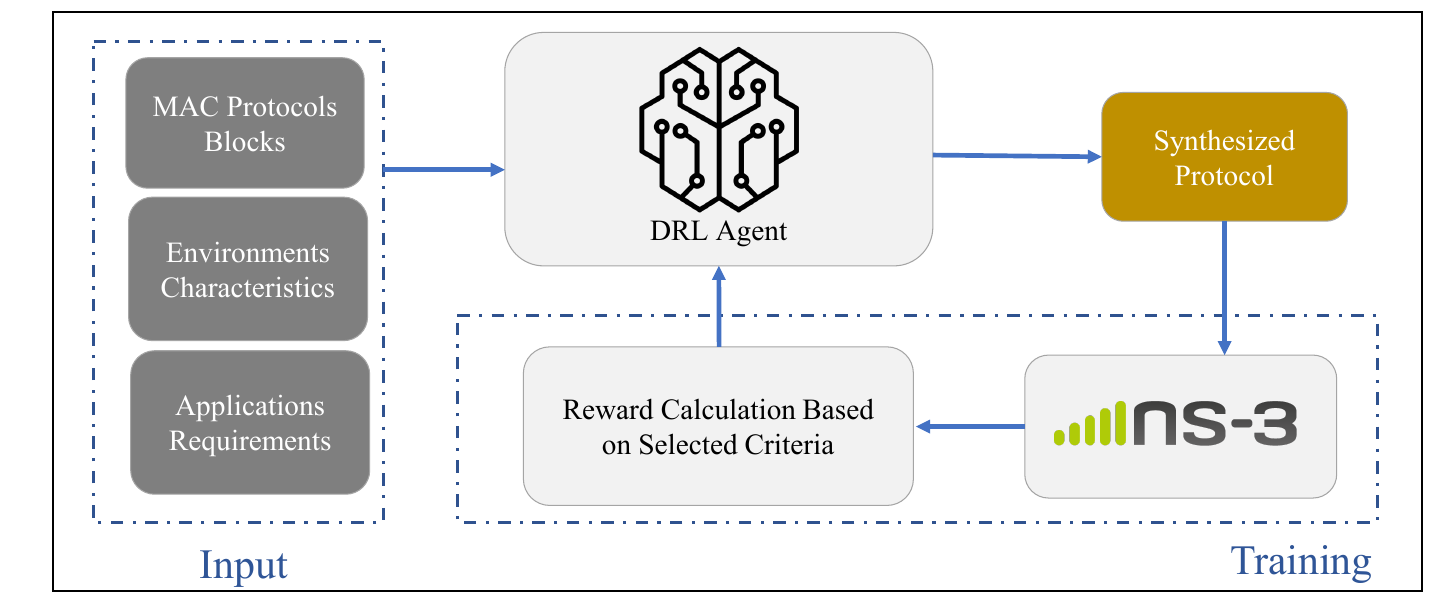}
 \vspace{-0.15cm}
	\caption{System concept.}
	\label{fig:System_model1}
 \vspace{-0.16cm}
\end{figure}

Figure \ref{fig:System_model1} illustrates our system concept.
The synthesis of MAC protocols is performed from a set of atomic building blocks interconnected through Machine Learning (ML) driven policies. To drive this process, we supply the ML framework with atomic building block functions such as backoff, sensing, defer, modulation and coding scheme and all of their related parameters along with application requirements and environment characteristics such as type of traffic and packet arrival rates, and number of nodes. The ML framework synthesizes a new protocol, which is subsequently evaluated in the \mbox{ns-3} simulator using a pre-defined reward, which is selected to fit the application requirements. Upon computation of the reward, feedback loops back to the ML framework. This iterative process continues until a protocol is constructed that incorporates the optimal set of building blocks for the current network/environment configuration. This modular approach empowers the agents to discern the most effective combination of these blocks, thereby generating novel MAC protocols or refining existing ones.

\begin{figure}[t]
	\centering
\includegraphics[width=0.47\textwidth,trim = 65mm 98mm 50mm 62mm,clip]{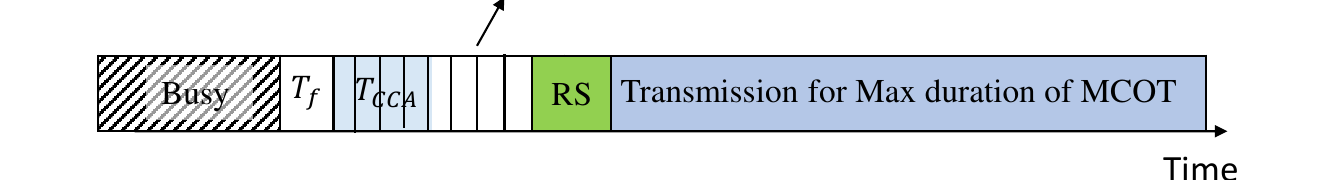}
  \vspace{-0.13cm}
 \caption{NR-U LBT4 Channel access mechanism.}
	\label{fig:NR-U}
 \vspace{-0.65cm}
\end{figure}

In this letter, we showcase the MAC protocols synthesis with our MADRL framework in a network comprising of 5G New Radio-Unlicensed (5G \mbox{NR-U}) gNBs.
5G \mbox{NR-U} is a radio access technology that is being developed by 3GPP and first introduced in Release 16\cite{release_16}. 
One of the key features is the channel access mechanism, namely Listen Before Talk (LBT), inherited from the LTE  licensed-assisted access (LTE-LAA), operating in 5 GHz unlicensed band. Before transmission, \mbox{5G NR-U} devices have to sense the channel to ensure harmonious coexistence with other unlicensed devices, such as IEEE 802.11ax. 
\mbox{5G NR-U} devices perform LBT procedure in the downlink, similar to Wi-Fi’s CSMA/CA protocol. As shown in Figure~\ref{fig:NR-U} after a period of idle channel, $T_f$, lasting $16~\mu s$, the gNB initiates Clear Channel Assessment (CCA) in a sequence of $d_i$ consecutive observation CCA slots, each with a duration of $9~\mu$. Subsequently, the deferred period $T_{df}$ is computed as $T_f$ + $d_i \times T_{CCA}$. Where $d_i$ is determined based on the priority assigned to various traffic types in the standard. If the channel remains idle throughout the defer time, the gNB initiates the backoff procedure by randomly selecting a number from the set $\{0,1,..., CW-1\}$ while continuing to sense the channel and decrementing the backoff counter. Upon reaching zero, the gNB starts transmission. If the channel becomes occupied during any of these slots, the backoff counter freezes, and the process restarts once the channel becomes idle again with the remaining backoff counter from the previous attempt. Upon gaining access to the channel, the gNBs can occupy it for a maximum duration, known as Maximum Channel Occupancy Time (MCOT). MCOT has different values for different traffic types as defined in the standard specification\cite{release_16}.
\section{Multi-Agent Deep Reinforcement Learning for Reconfigurable MAC protocol}
Our goal is to develop a multi-agent learning algorithm for reconfigurable MAC that can effectively adjust according to real wireless network scenarios while overcoming the complexity due to large parameter spaces and partial observability of the environment. First, we define the agent's decision process as a Partially Observable Markov Decision Process (POMDP), consisting of observations, actions, and rewards. Thereafter, we leverage the proximal policy optimization (PPO) algorithm to train the agents efficiently in a distributed manner.
\subsection{Problem Formulation}
Our ultimate objective is to maximize the long-term throughput averaged over all $N$ gNBs and episode time steps, $T$ as follows:
\vspace{-1.8mm}
\begin{equation} \label{eqn0}
\max \frac{1}{T} \frac{1}{N} \sum_{t=1}^{T} \sum_{j=1}^{N} {Th}_j(t),
 \vspace{-1.6mm}
\end{equation}
where $Th$ is the aggregated downlink throughput per gNB. 
We consider a network environment where multiple gNBs are deployed, each serving a particular area with a diverse set of traffic types, including Poisson traffic with different arrival rates, $\lambda$ and augmented and/or virtual reality (AR/VR) traffic, modeled as bursty traffic with different frame rates \cite{Zorzi1}. We assume a partial observability of the environment at each gNB. Each node has complete autonomy in creating the MAC protocol and adjusting the protocol parameters. This means each agent has the capability to manipulate the deferred period ($T_{df}$) by modifying the size and number of clear channel assessment slots or the defer time $d_f$, as well as adjusting the backoff number, size, and its functions. This allows each agent to create various types of MAC protocols. Moreover, agents can control parameters such as the energy detection threshold ($ED_{Th}$) and the transmission power ($Tx$). 
\subsection{Partially Observable Markov Decision Process (POMDP)}
\textbf{Observation space $\mathcal{O}$}: Observation ($\mathcal{O}_x$) = $\langle$$Current Action_x$, $NN_x$, $RSSI_C$, $RSSI_I$, $Throughput_x$, $TR_x$, $Delay_x$, $Airtime_x$$\rangle$ The observation space of agent $x$ is defined as tuple, which includes the current MAC protocol blocks specified by the $Current Action_x$, the number of visible nodes in the surrounding area $NN_x$, which depends on the energy detection threshold, the interference from other nodes ($RSSI_I$), and the received power level from the connected user ($RSSI_C$). The agent can also calculate the throughput, $Throughput_x$ and the delay $Delay_x$. The parameter $Airtime_x$ is the airtime occupied by other users on the channel, which is obtained through the sensing capability of the agent. We assume that each node broadcasts its traffic characteristics $TR_x$ and the aggregated downlink throughput. Broadcasting can be done using the X2 interface defined in NR protocols for communication between neighbour nodes.

\textbf{Action $\mathcal{A}$:}
Action ($\mathcal{A}_x$) = $\langle$$MCOT_X$, $T_x$, $MCS_x$, $ED_{Th,x}$, $T_{df}$, $Backoff\_type_x$, $CW\_min_x$, $Sensing~slot~duration_x$$\rangle$
The action space $\mathcal{A}_x$ for each agent $x$ is defined as a tuple containing MAC block functions and their parameters that determine the behaviour of the MAC protocol. These parameters include the backoff function type $Backoff\_type_x$ and its relevant parameters, such as the sensing slot duration $Sensing~slot~duration_x$, the minimum contention window size $CW\_min_x$, the energy detection threshold $ED_{Th,x}$ and the defer time $T_{df}$. Additionally, the tuple specifies the modulation and coding scheme $MCS_x$, the maximum channel occupancy time $MCOT_X$, and the transmission power $T_x$. Each agent makes decisions on whether to include specific MAC protocol blocks and choose appropriate values for parameters. Table \ref{tab:parameters} provides a summary of the action space parameters and their corresponding values. Each parameter in the action space has a range of possible values, allowing agents to make diverse decisions when configuring the MAC protocol.

\begin{figure}[t]
\centering
\includegraphics[width=0.40\textwidth,trim = 65mm 10mm 50mm 62mm,clip]{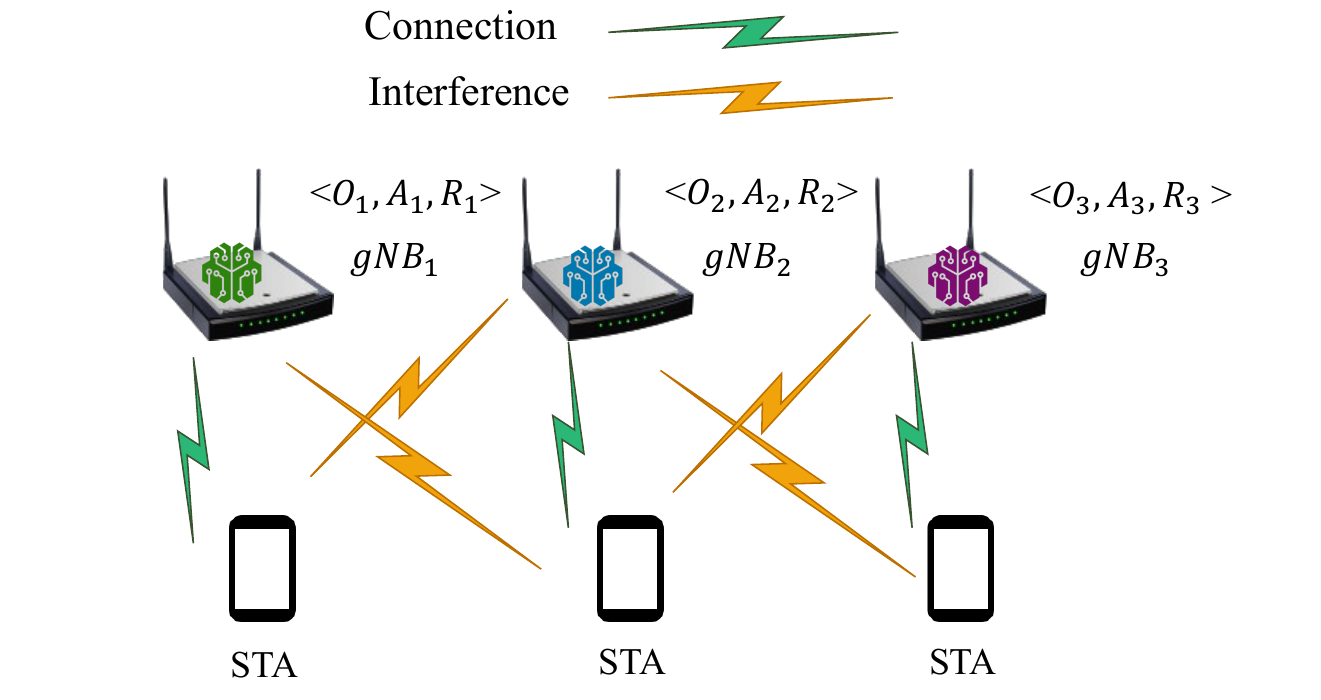}
\vspace{-0.22cm}
\caption{Distributed training and execution architecture.}
	
	\label{Distributed_2}
\end{figure}
\textbf{Reward $\mathcal{R}$:}
Each agent broadcasts its throughput and traffic rate to the nodes within its range. Each node can also calculate the airtime of other nodes within its range. We define the reward for each agent as follows:
 \begin{equation}
 \mathcal{R}_i = \frac{\overline{Th}_i}{\overline{\lambda}_i} - \alpha\overline{t}_{air,i} ,
 \vspace{-1.6mm}
 \end{equation}
 where $\overline{Th}_i$ represents the mean normalized aggregated downlink throughput of $i_{th}$ network and $\overline{\lambda}_i$ is the normalized traffic arrival rate and $\overline{t}_{air,i}$ denotes the normalized airtime of $i_{th}$ gNB. Both $\overline{\lambda}_i$ and $\overline{t}_{air,i}$ are normalized with respect to other nodes within their sensing range.  The reward encourages effective usage of the channel by minimizing airtime while maximizing throughput. Additionally, the reward function discourages greediness among agents by considering the throughput and airtime of other nodes within range.

\begin{table}[h!]
\small
 \vspace{-0.05cm}
	\caption{The Action Space}
  \vspace{-0.144cm}
	\begin{tabular}{|p{0.1cm}|p{2.4cm}|p{2.36cm}|p{2.41cm}|}
		\hline
		\textbf{} &  \textbf{Action parameter} & \textbf{Values Range}& \textbf{Standard value} \\
		\hline
		$a_1$& Sensing Slot Size
        &\{0, 1, 2, \ldots, 20\}&9
		\\
		\hline
		$a_2$& Backoff type
		& Off, EDID,& BEB\\& &BEB, Constant& \\
		\hline
		$a_3$&  Minimum CW
		& \{0, 1, 2, \ldots, 63\}& 15
        \\
		\hline
		$a_4$
		  & MCOT
        & \{0, 1, 2, \ldots, 10\} & 2, 3, 5, 8
		\\
		\hline
	    $a_5$& MCS 
        & \{0, 1, 2, \ldots, 28\}& Auto. Rate Control
        \\ 
		\hline
        $a_{6}$& $T_{df}$ 
        & \{0, 1, 2, \ldots, 20\}&16
        \\
		\hline
        $a_{7}$& $ED_{Th}$ [dBm] 
        & \{-90, -89, \ldots, -60\} &-62 dBm
        \\
        \hline
        $a_{8}$&  $T_x$  [dBm]
        & \{10, 11, \ldots, 30\}&23 dBm
        \\
		\hline
	\end{tabular}
	\label{tab:parameters}
  \vspace{-0.12cm}
\end{table}

\begin{algorithm}[t]
\small
\hspace*{\algorithmicindent} \textbf{Input:} 
 {All training parameters from Table \ref{tab:Training Parameter}\\}
 \hspace*{\algorithmicindent} \textbf{Output:}
 {$\pi_{\Theta_x}$, $V_{\phi_x}$} where $\forall x\in \{1,2,...,NN\}$
	\begin{algorithmic}[1]
    \State $\forall x\in \{1,2,...,NN\}$ Initialize the actor $\pi_\Theta(a|s)$ and critic $V_\phi$ with random parameters $\Theta_x$ and $\phi_x$, respectively. 
	\For{\mbox{iteration=1, 2,...$N_{eps}$}}
    \State Initialize the environment
    \State $j=0$
    \While{$ j< T$}
    \For{\mbox{Agents at $gNB_x$with $x\in\{1,2,...,NN\}$} are deployed in parallel}
    \State {Generate an experience set of $N$ time steps by following MAC block policy $\pi_\Theta$ from every agent in parallel
    \State Collecting $(o^x_t-1,a^x_t-1,r^x_t,o^x_t)$}
    \State {For each step calculate advantage function $A_t=\sum_{i=t}^L\gamma^{i-k}r_i-V_\phi(o_t)$ and return function $G_t=A_t+V_\phi(o_t)$}
    \State {Each agent Collect a subset of $M$ random sample (mini-batches) from the current set of experience separately}
    \State {Calculate value function loss  $L^{VF}=\frac{1}{2M}\sum_{i=1}^M(G_i-V_\phi(s_i))^2$ and Minimize the $L^{VF}$ using gradient descent and update $\Phi$}
    \State {Compute ratio $r_t(\theta)$and entropy loss $S[\pi(a_t|o_t)]$} and calculate surrogate objective $L^{\mathrm{CLIP}}=\frac{1}{M}\sum_{i=1}^M[-\min(r(\Theta) A, \mathrm{clip}(r, 1 - \epsilon, 1 + \epsilon) A)+cS[\pi_\Theta(a_t|o_t)]$
    \State {Update the policy network parameters $\theta$ by maximizing $L^{\mathrm{CLIP}}$, taking gradient ascent}
    \State $j=j+N$
     \EndFor 
    \EndWhile
     \EndFor 
		\caption{Multi-Agent Distributed Training and Distributed Execution (DTDE) for Reconfigurable MAC protocol}
		\label{DLDI_Algorithm}
	\end{algorithmic}
\end{algorithm}
\newlength{\textfloatsepsave}
\setlength{\textfloatsepsave}{\textfloatsep}
\setlength{\textfloatsep}{0pt}
\vspace{-8mm}
\subsection{PPO for multi-agent reconfigurable wireless MAC protocol}
\label{PPO}
We use Proximal Policy Optimization (PPO) for designing MAC policies across different deployment scenarios \cite{schulman2017proximal}, developed by OpenAI. PPO is an actor-critic algorithm, meaning it employs two separate neural networks, for value and policy estimation. We adopt a fully distributed approach for both learning and execution. Each gNB node hosts a single agent dedicated to training and inference tasks as illustrated in Figure \ref{Distributed_2}. Each agent operates autonomously with its own dedicated neural networks, ensuring complete autonomy and decentralization.
As shown in Algorithm 1, every agent initializes the value networks $V_\phi$ and policy networks $\pi_\Theta$ with the respective parameters $\phi$ and $\Theta$ and maintains separate mini-batches. For each iteration, the policy at each node is executed in the environment independently, and each agent accumulates its experiences according to the PPO algorithm (lines 5-6), facilitating individualized learning and adaptation.
Subsequently, the advantage function is computed for each time step at each node. The advantage function $A_t$ measures the potential benefit of choosing a particular action in a certain state compared to the average outcome expected when following the current policy, and is defined as follows:
\vspace{-1mm}
\begin{equation} \label{eq:PPO22}
A_t=\sum_{k=0}^{T-t-1}\gamma^{k}r_i-V_\phi(o_t),
 \vspace{-1.6mm}
\end{equation}
where the first term, discounted returns $G_t=\gamma^{k}r_i$, is calculated using the collected rewards, and $V_\phi(o_t)$ is the value estimate for each observation $o_t$ from the value network.

To optimize the policy and value networks, we randomly collect samples and add them to mini-batches. The value network is optimized by minimizing the value loss, which is defined as the mean squared error between the predicted values and the computed target values.
\vspace{-2mm}
\begin{equation}
\label{eq:PPO23}
L^{VF}=\frac{1}{2M}\sum_{i=1}^M(G_i-V_\phi(o_i))^2
 \vspace{-1.6mm}
\end{equation}
Following this, we proceed to update the parameters, $\phi$, of the value network. This is accomplished by minimizing the value loss by using gradient descent.
Concurrently, the optimization process in PPO involves updating the policy network by maximizing a clipped surrogate objective, which is given by$L^{\mathrm{CLIP}} = \frac{1}{M}\sum_{i=1}^M[-\min(r(\Theta) A, \mathrm{clip}(r, 1 - \epsilon, 1 + \epsilon) A) \nonumber + cS[\pi(a_t|o_t)].$ 
This loss function uses the ratio of the new policy to the old policy, which is computed as follows: 
\begin{equation} \label{eq:PPO25}
r_t(\theta)=\frac{\pi_\theta(a_t|o_t)}{\pi_{\theta_{old}}(a_t|o_t)}
 \vspace{-1.6mm}
\end{equation}
Additionally, a clipping function is used to ensure that updates are not extreme, as significant deviations could destabilize learning progress. Here, $\epsilon$ represents the clipping parameter.
The entropy term, denoted by $S[\pi(.|s;\Theta)]$, encourages exploration by the policy and prevents premature settling on suboptimal deterministic policies. The entropy term is defined as:
\vspace{-4mm}
 \begin{equation} \label{eq:PPO33}
     S[\pi_\Theta(a_t|o_t)]=-\sum_a\pi_\Theta(a_t|o_t)\log \pi_\Theta(a_t|o_t),
       \vspace{-1.6mm}
\end{equation}
where parameter $c$ acts as a coefficient that controls the weight of the entropy term. 
Ultimately, the policy network parameters $\theta$ are updated by maximizing the objective using a gradient ascent.
\begin{figure}[t]

	\centering
\includegraphics[width=0.40\textwidth,trim = 70mm 25mm 70mm 55mm,clip]{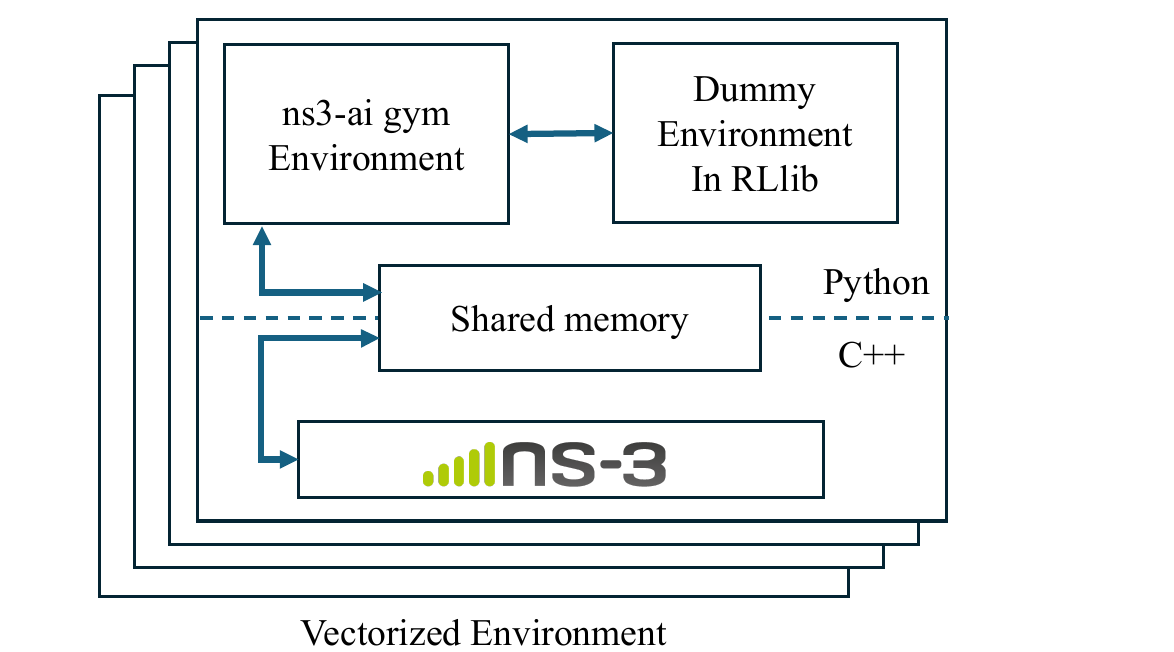}
 \vspace{-0.10cm}
    \caption{ Integration of ns3-ai gym with RLlib using a dummy environment.}
	\label{fig:Learning_rate}
\end{figure}
During the training phase, the broadcasting of rewards and traffic characteristics can be accomplished using the X2 interface defined in NR protocols for communication between neighbouring nodes or added to the packet header. 
\vspace{-2mm}
\section{Simulation and Learning Environment}
We implement our MADRL framework by integrating ns3-ai and RLlib. To ensure compatibility between ns3-ai and RLlib, we have created a dummy environment within RLlib, which collects the data from the ns3-ai gym environment. Our primary simulation is the \mbox{ns-3}, with ns3-ai responsible for transferring actions and observations between \mbox{ns-3} and RLlib's dummy environment, as illustrated in Figure~\ref {fig:Learning_rate}. Observations collected from the \mbox{ns-3} environment are relayed to the dummy environment, where agents analyze them to determine suitable actions. These actions are then directly applied to the agent within the \mbox{ns-3} simulation environment, enabling uninterrupted simulation. We use the 5G \mbox{NR-U} module, which is a full-stack implementation of NR, including the channel access mechanism specified for NR-U technology. We have ensured its full functionality and compatibility with ns3-ai.

\begin{figure}
	\centering
\includegraphics[width=0.39\textwidth,trim = 0mm 0mm 0mm 0mm, clip]{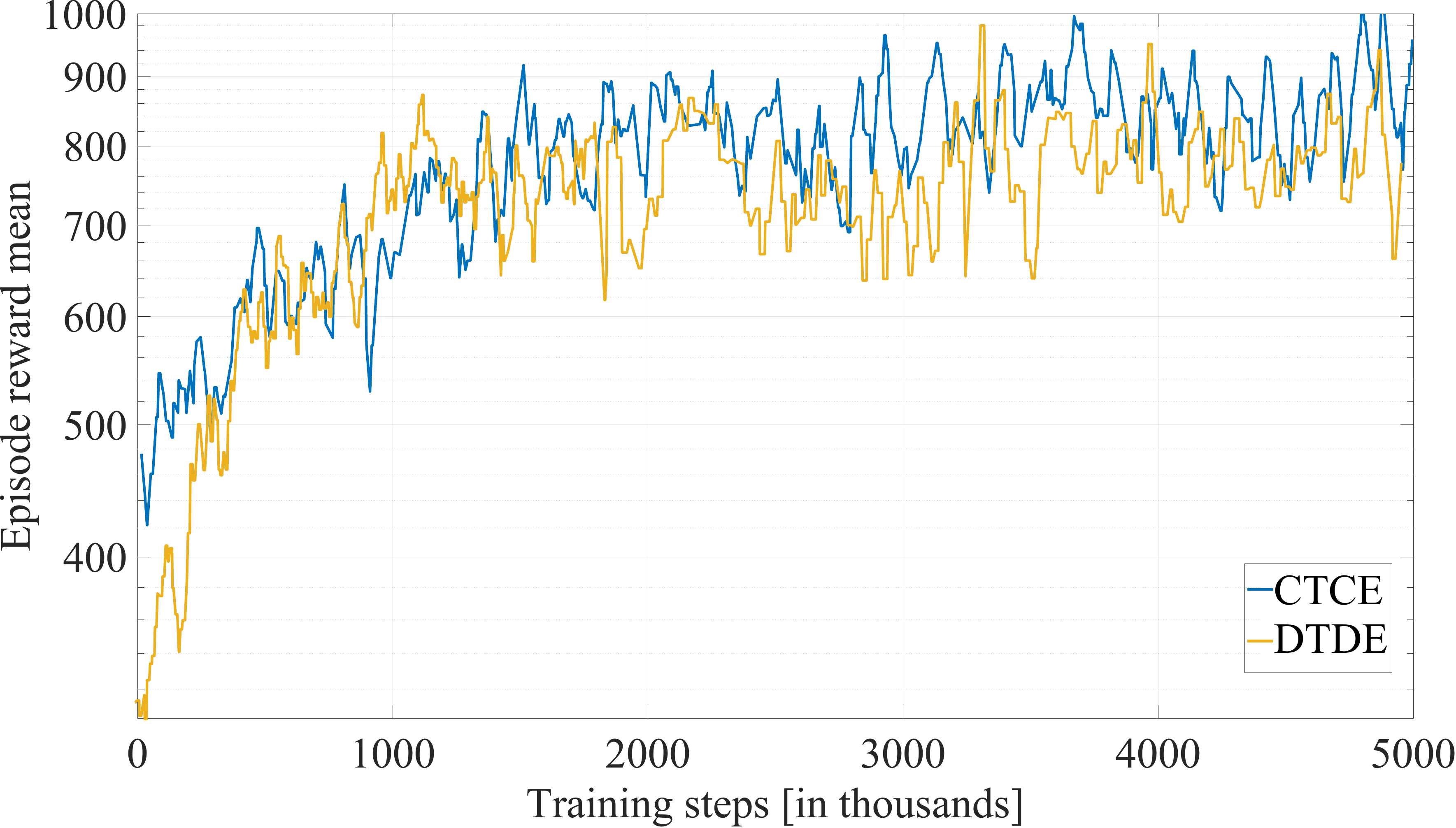}
\vspace{-0.10cm}
	\caption{Learning curves comparing the convergence of proposed DTDE against the centralized approach CTCE.}
	\label{fig:learning_curve}
\end{figure}

The simulation and training processes were conducted on a server equipped with 2 GPU units and 64 CPU cores. 
The \mbox{ns-3} simulations ran on the CPU, while the training process, involving machine learning algorithms and neural network models, ran simultaneously on the GPU, for a faster convergence of the overall learning process. 
Figure~\ref{fig:learning_curve} illustrates the learning convergence of the proposed distributed training and execution approach, DTDE, against the centralized training and execution approach, CTCE introduced in \cite{keshtiarast}. In CTCE, a single agent is responsible for both training and execution. The distributed approach converges notably faster than the centralized approach, which can be attributed to the CTCE’s requirement for the agent to manage a significantly larger action space, thereby slowing down the learning process. It is also worth noting that due to the distributed nature of the system and the lack of full control and knowledge over other nodes, DTDE achieves a slightly lower mean reward compared to centralized learning.

\begin{figure*}[!ht]
  
    \centering
    \subfigure[{Low-density traffic}]{
    \hspace{0.5mm}
    \vspace{-1mm}
    \includegraphics[width=0.220\textwidth,trim = 0mm 0mm 0mm 0mm,clip]{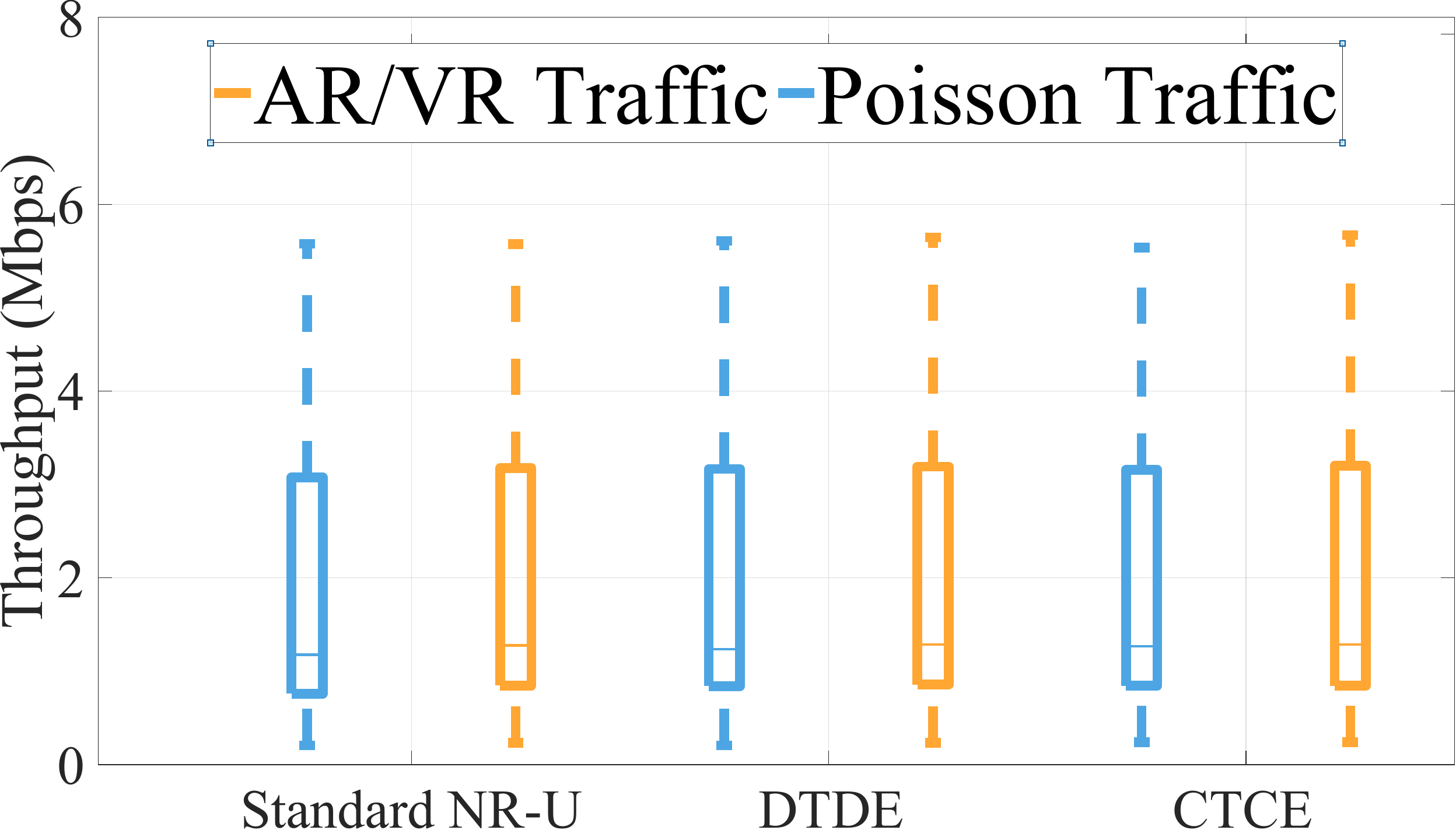}
    \label{TSUS_1_1}
    }
    \subfigure[{Medium-density traffic}]{
\hspace{-1mm}\includegraphics[width=0.235\textwidth,trim =0mm 0mm 0mm 0mm,clip]{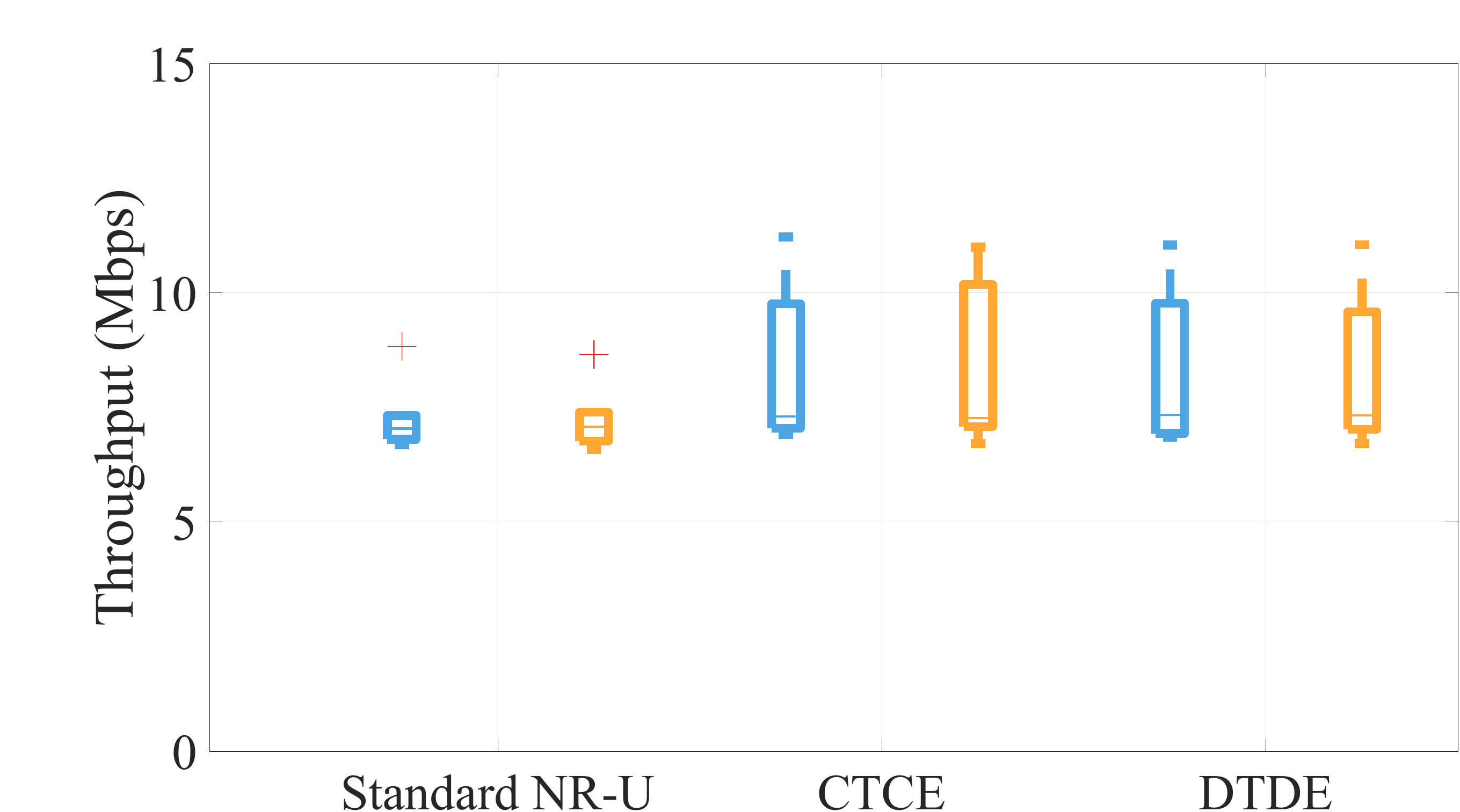}
    \label{TSUS_1_2}
    }
    \subfigure[{High-density traffic}]{
    \includegraphics[width=0.234\textwidth,clip]{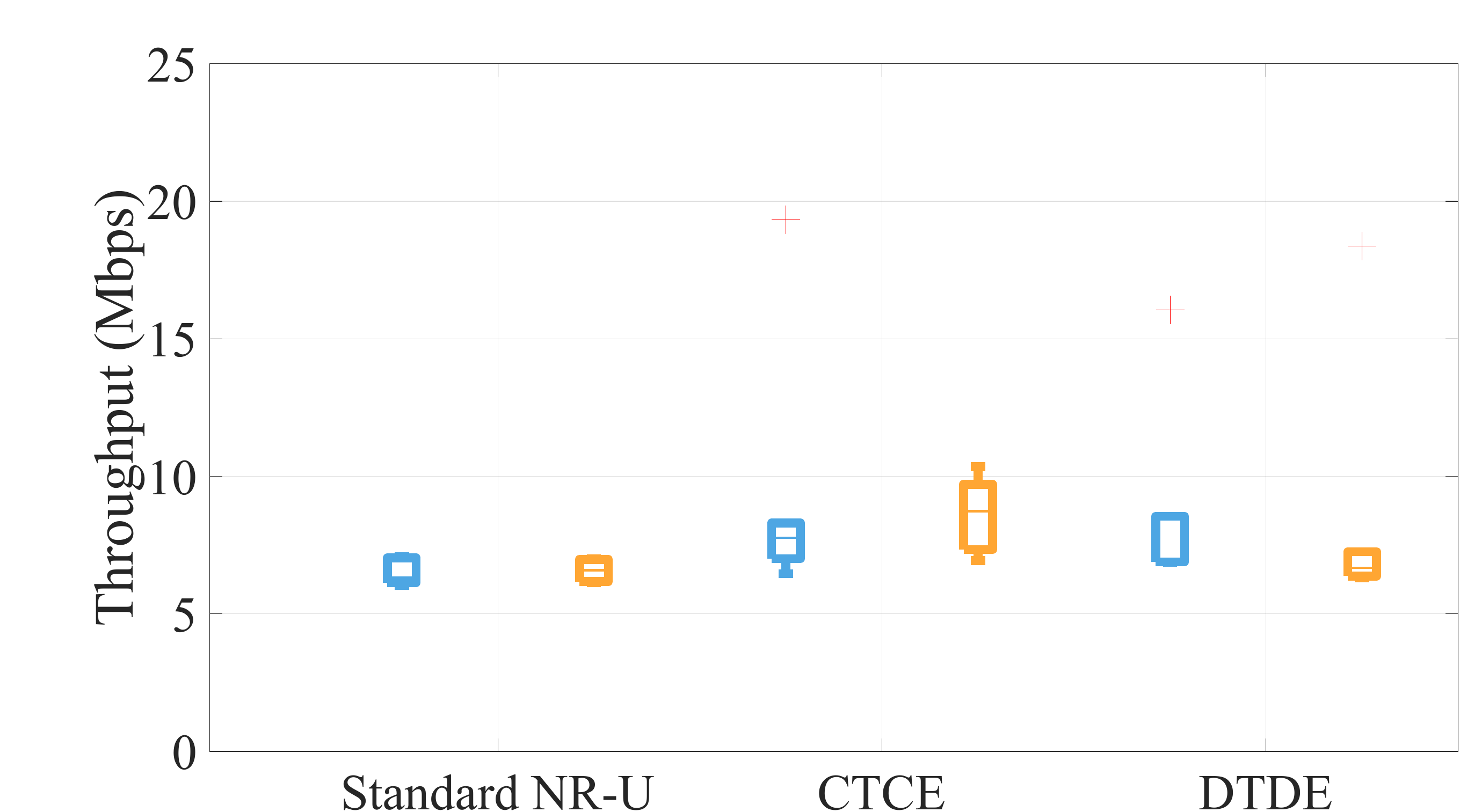}
    \label{TSUS_1_3}
    }
    \subfigure[{Random-density traffic}]{
    \includegraphics[width=0.232\textwidth,clip]{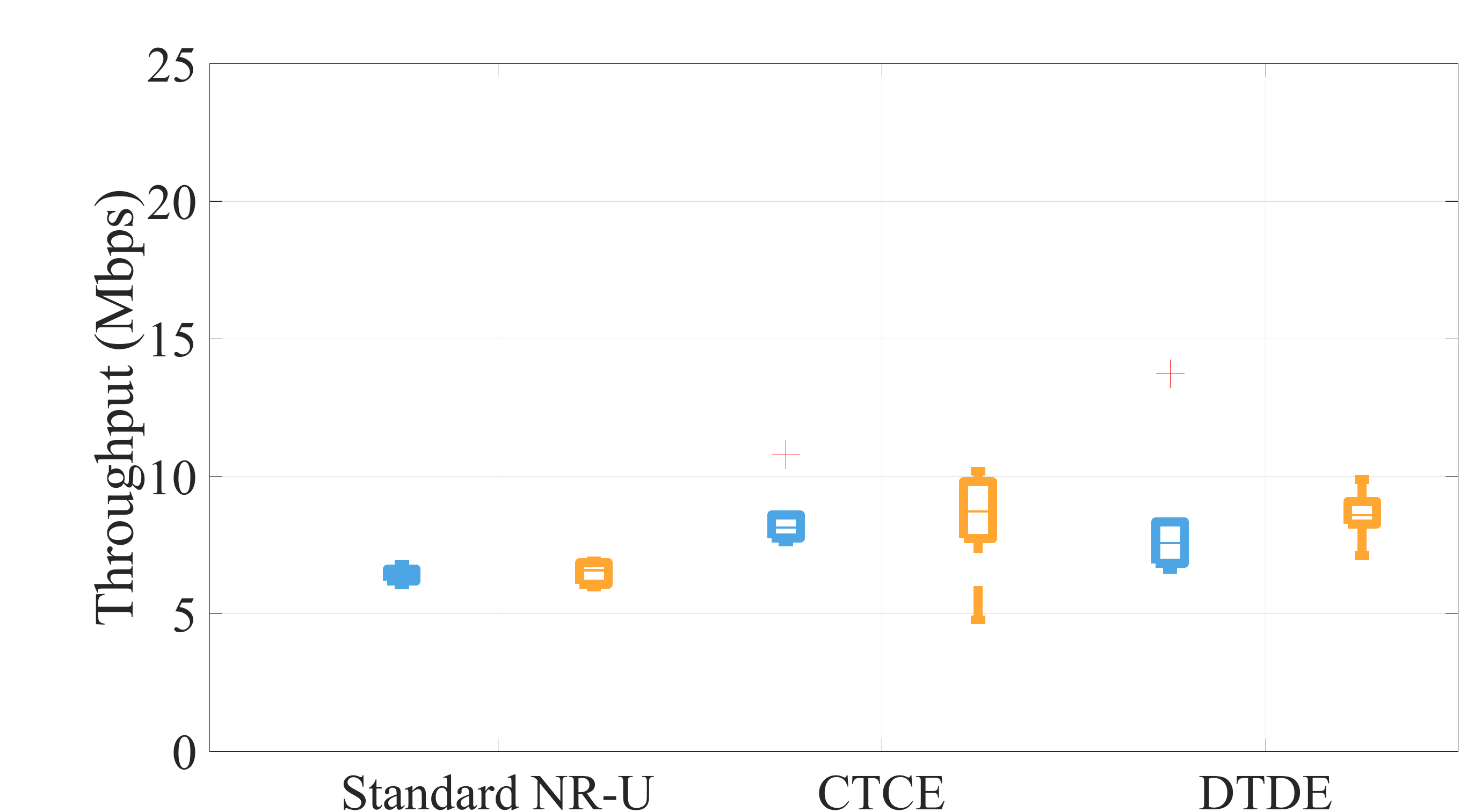}
    \label{TSUS_1_4}
    }
    \subfigure[{Low high-density traffic}]{
    \includegraphics[width=0.23\textwidth,trim = 15mm 0mm 0mm 0mm,clip]{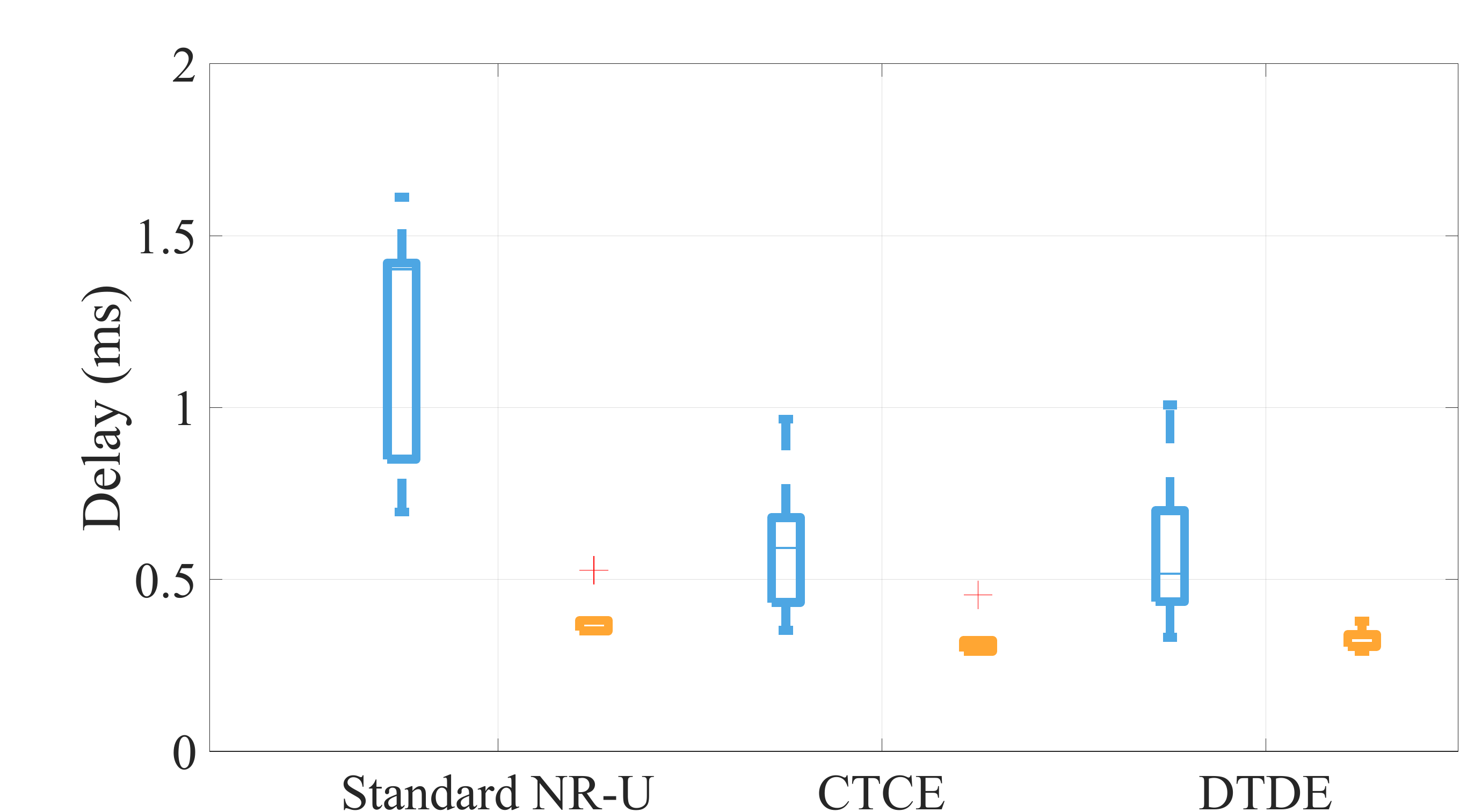}
    \label{TSUS_2_1}
    }
    \subfigure[{Medium high-density traffic}]{
    \includegraphics[width=0.23\textwidth,trim = 15mm 0mm 0mm 0mm,clip]{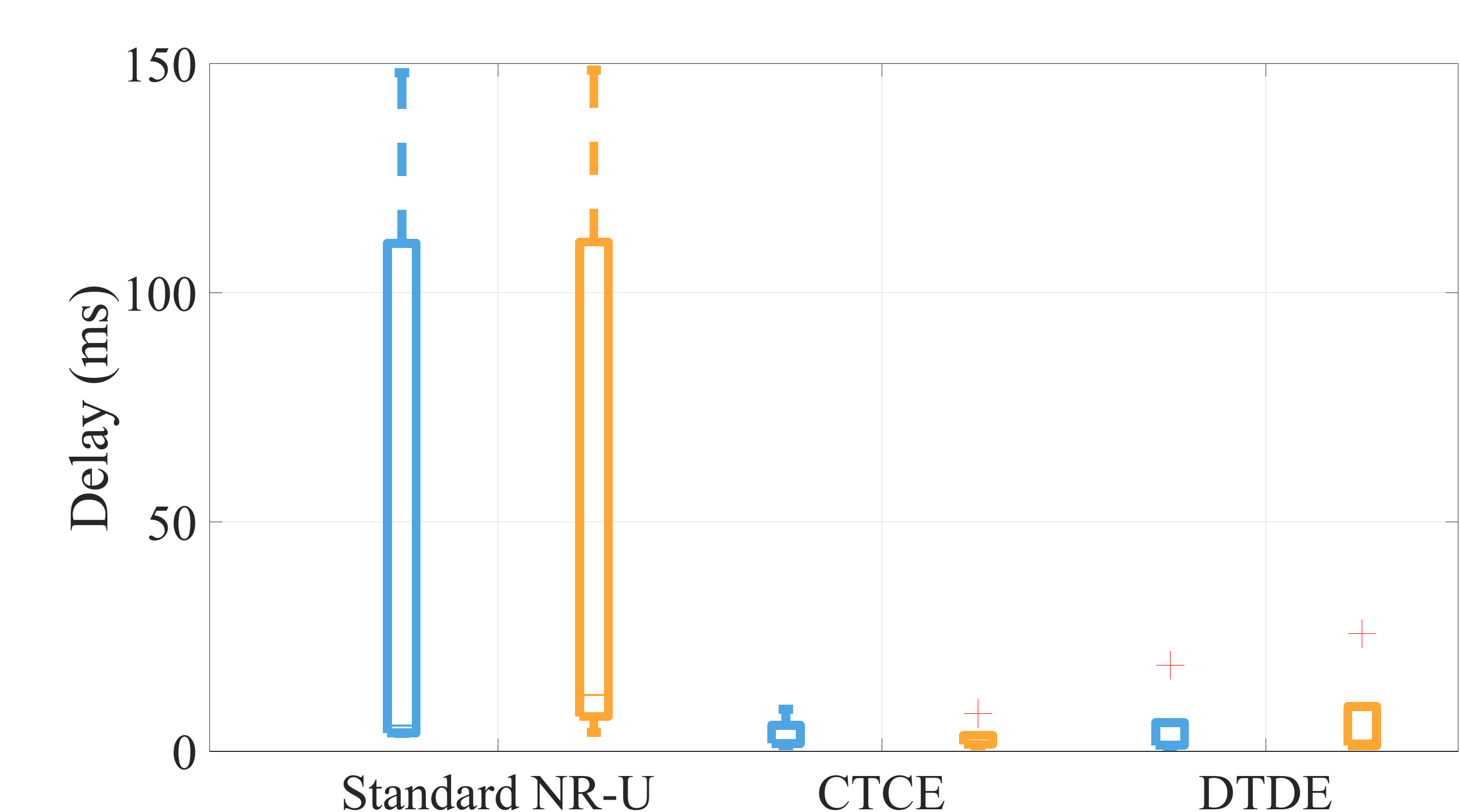}
    \label{TSUS_2_2}
    }
    \subfigure[{High-density traffic}]{
    \includegraphics[width=0.23\textwidth,trim = 15mm 0mm 0mm 0mm,,clip]{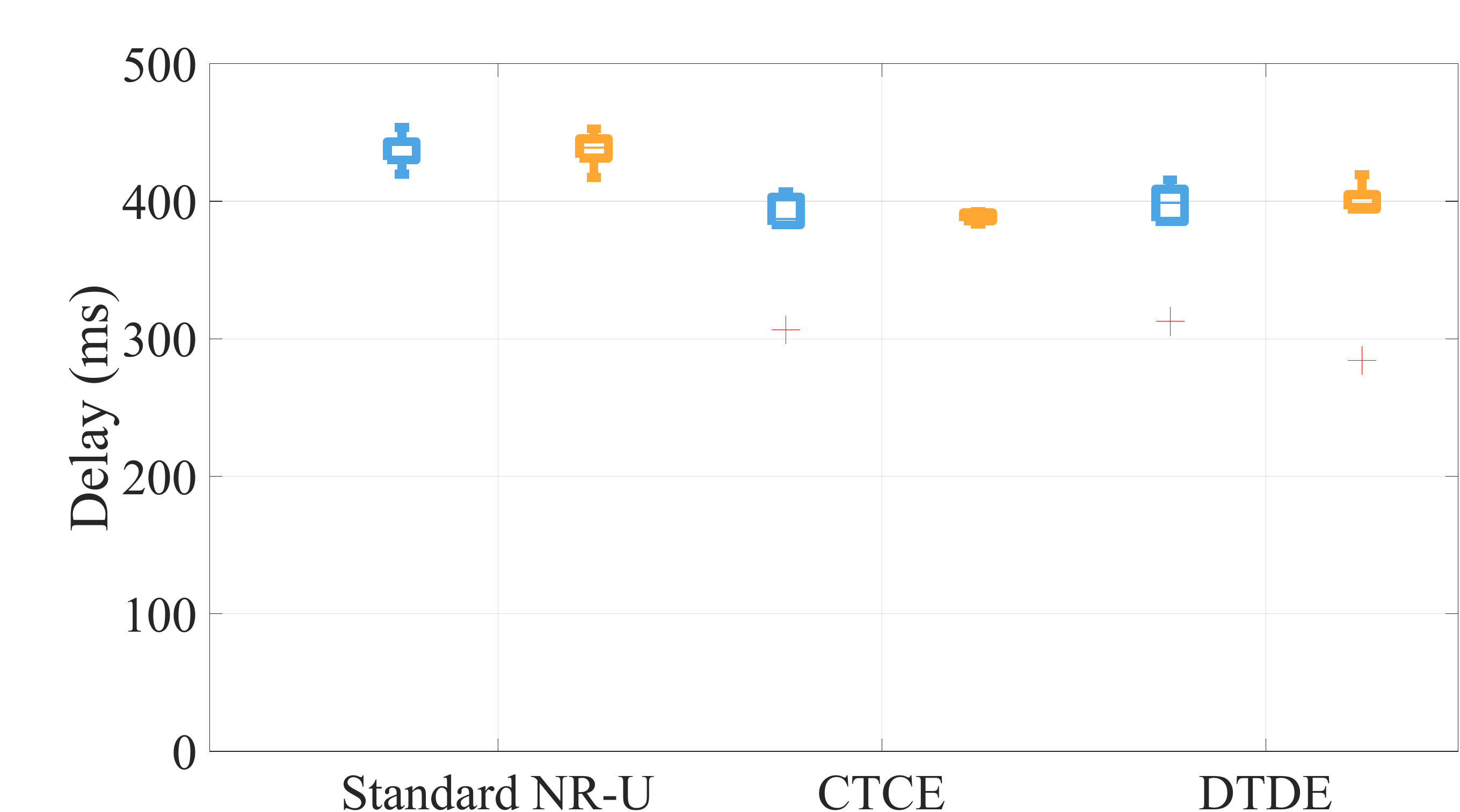}
    \label{TSUS_2_3}
    }
    \subfigure[{Random-density traffic}]{
    \includegraphics[width=0.23\textwidth,trim = 15mm 0mm 0mm 0mm,clip]{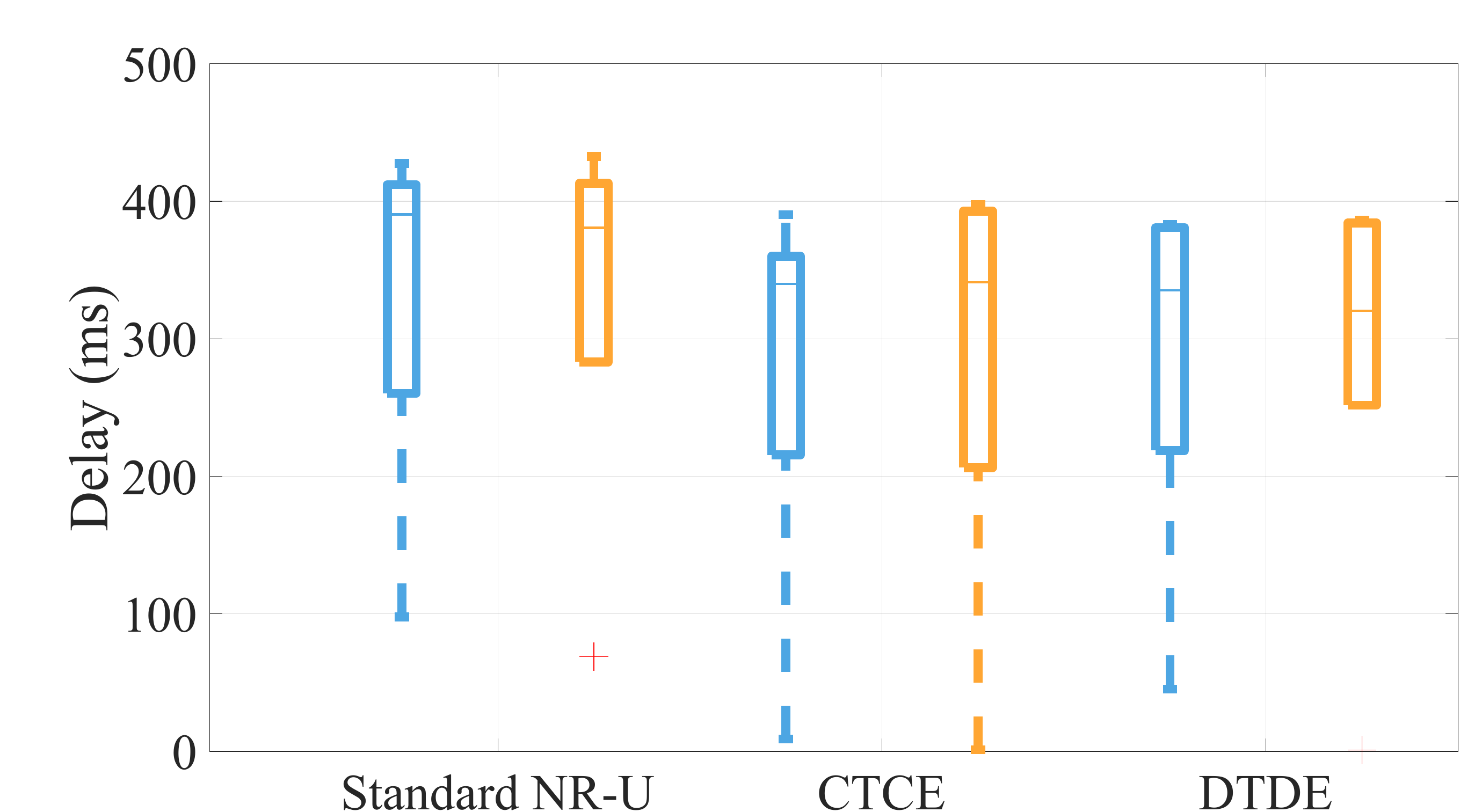}
    \label{TSUS_2_4}
    }
    \vspace{-0.25cm}
    \caption{Evaluation results comparing standard 5G \mbox{NR-U} with our distributed MADRL framework for various traffic densities.}
	\label{fig:results}
 \vspace{-0.48cm}
\end{figure*}

\begin{table}[ht]
\small
\centering
 \vspace{-0.13cm}
	\caption{Training and Environment Parameters}
  \vspace{-0.14cm}
	\begin{tabular}{|p{4.9cm}|p{2.9cm}|}
		\hline
		 Number of networks (NN) & 1-6\\
		\hline
		 Frequency & 6 GHz
		\\
		\hline
		Bandwidth & 20 MHz
		\\
        \hline
		Traffic characteristic (TR): Poisson and AR/VR with arrival rates $\lambda$ & $\lambda= $[\ 0 - 3000] \\  
		\hline
        Packet size &$1500$\\
        \hline
		\textbf \textbf{Learning Rate, Optimizer} & 0.001, Adam\\
		\hline
        
		Policy & RNN (2 layers of 256) 
        \\
		\hline
        batch size, $M$&1000
        \\
		\hline
        Step size  & 0.1 s
        \\
        \hline
        Episode duration  & 50 s
        \\
        \hline
        
        $\alpha$ & 0.3 \\ \hline
	\end{tabular}
	\label{tab:Training Parameter}
 \vspace{-0.45cm}
\end{table}
\section{Performance Evaluation}
For the evaluation of our distributed multi-agent approach, i.e., DTDE, we randomly deployed six 5G \mbox{NR-U} gNBs in an area of $200\times200$ $m^2$. Each gNB is connected to at least one UE. At each gNB a single agent is deployed that learns the optimal MAC blocks and their parameters based on the current environment and application requirements. We consider Poison and AR/VR traffic, which has bursty characteristics \cite{Zorzi1}. We evaluate the performance of the synthesized MAC protocol in terms of mean downlink throughput and average end-to-end packet delay per gNB.

Figure~\ref{fig:results} shows the evaluation results for various traffic densities and types. We consider four traffic densities based on packet arrival rates $\lambda$, i.e., low traffic (10 to 500 packets/sec), medium traffic (500 to 1000 packets/sec), high traffic (1000 to 3000 packets/sec), and random-rate traffic (10 to 3000 packets/sec). We ran our distributed approach for 10 episodes, with each agent using its learning model. Afterwards, we compared its performance with that of the standard-based 5G NR-U and the centralized approach, i.e., CTCE. The system parameters of 5G \mbox{NR-U} are listed in Table \ref{tab:parameters}. 


Figures \ref{TSUS_1_1} and \ref{TSUS_2_1} illustrate the distribution of mean throughput and delay across all nodes in the environment for low-density Poisson and AR/VR traffic. As the contention for channel access is minimal and all the packets in the queue can be successfully transmitted, both baselines (the standard 5G \mbox{NR-U} and CTCE), and our distributed multi-agent NR-U protocol show similar throughput. However, in terms of delay, both learning approaches show improvement due to the selection of appropriate MAC blocks and parameters and removing the unnecessary overhead in the standards 5G \mbox{NR-U}.

Figures \ref{TSUS_1_2}, \ref{TSUS_1_3}, and \ref{TSUS_1_4} display the throughput distributions for medium, high, and mixed-rate traffic. The results demonstrate that our distributed approach improves mean throughput by at least $10\%$, primarily due to the reduction of carrier sensing overhead and the dynamic selection of MAC protocol blocks tailored to each node's specific requirements. This is achieved by selecting the appropriate backoff algorithm, deferring, and sensing parameters at each node based on the environmental characteristics observed by each agent.

It is also worth noting that some nodes achieved significantly higher throughput without adversely affecting others, as indicated by the red outlier points. This improvement is due to our framework's ability to select not only the optimal MAC protocols for each scenario but also to adjust transmission power levels to minimize interference with other coexisting nodes. Additionally, nodes adjust their sensitivity to interference from neighbouring nodes by changing $ED_{Th}$, which allows nodes to access the channel more freely, similar to the Basic Service Set (BSS) coloring technique used in Wi-Fi technology.
As a result, nodes gain more opportunities to transmit, leading to a significant reduction in end-to-end packet delay, as shown in Figures \ref{TSUS_2_2}, \ref{TSUS_2_3}, and \ref{TSUS_2_4}. 

Overall, our distributed multi-agent NR-U protocol consistently surpasses the standard 5G NR-U protocol and closely matches the performance of the centralized baseline, despite each agent having only partial observation compared to the centralized model, which possesses complete knowledge. This success is largely attributed to our reward function, which ensures that each agent at each gNB considers not only its own performance but also that of neighboring nodes within its sensing range.

\vspace{-1mm}
\section{Conclusions}
In this letter, we have proposed a MADRL framework that leverages distributed multi-agent machine learning to empower individual network nodes to autonomously optimize design and configure MAC protocols, thus overcoming the limitations of centralized decision-making. By enabling nodes to customize their Medium access based on local observations, our approach offers adaptability and scalability tailored to specific environmental conditions. Through extensive simulations, we have demonstrated the superiority of MADRL-synthesized protocols over the legacy 5G \mbox{NR-U} MAC, highlighting the potential of the new protocol design approach to enhance QoS for future wireless applications.
\bibliographystyle{IEEEtran}
\bibliography{Letter_recmac}
\end{document}